\documentclass[aip,rsi,amsmath,amssymb,preprint]{revtex4-1} 
\usepackage{graphicx}
\usepackage{dcolumn}
\usepackage{gensymb}
\usepackage{color,soul}	
\usepackage[noabbrev]{cleveref}

\begin{document}


\title{An integrated and multi-purpose microscope for the characterization of atomically thin optoelectronic devices}

\author{Adolfo De Sanctis}
\author{Gareth F. Jones}
\author{Nicola J. Townsend}
\author{Monica F. Craciun}
\author{Saverio Russo}
\email[Correspondence should be addressed to: ]{S.Russo@exeter.ac.uk}
\affiliation{Centre for Graphene Science, College of Engineering, Mathematics and Physical Sciences, University of Exeter, Exeter EX4 4QF, United Kingdom}

\date{\today}

\begin{abstract}
Optoelectronic devices based on graphene and other two-dimensional (2D) materials, such as transition metal dichalcogenides (TMDs) are the focus of wide research interest. They can be the key to improving bandwidths in telecommunications, capacity in data storage, new features in consumer electronics, safety devices and medical equipment. The characterization these emerging atomically thin materials and devices strongly relies on a set of measurements involving both optical and electronic instrumentation ranging from scanning photocurrent mapping to Raman and photoluminescence (PL) spectroscopy. Furthermore, proof-of-concept devices are usually fabricated from micro-meter size flakes, requiring microscopy techniques to characterize them. Current state-of-the-art commercial instruments offer the ability to characterize individual properties of these materials with no option for the \textit{in situ} characterization of a wide enough range of complementary optical and electrical properties. Presently, the requirement to switch atomically-thin materials from one system to another often radically affects the properties of these uniquely sensitive materials through atmospheric contamination. Here, we present an integrated, multi-purpose instrument dedicated to the optical and electrical characterization of devices based on 2D materials which is able to perform low frequency electrical measurements, scanning photocurrent mapping, Raman, absorption and PL spectroscopy in one single set-up with full control over the polarization and wavelength of light. We characterize this apparatus by performing multiple measurements on graphene, transition metal dichalcogenides (TMDs) and Si. The performance and resolution of each individual measurement technique is found to be equivalent to that of commercially available instruments. Contrary to nowadays commercial systems, a significant advantage of the developed instrument is that for the first time the integration of a wide range of complementary opto-electronic and spectroscopy characterization techniques is demonstrated in a single compact unit. Our design offers a versatile solution to face the challenges imposed by the advent of atomically-thin materials in optoelectronic devices.
\end{abstract}


\maketitle 

\section{Introduction} \label{sec:Intro}
Since the discovery of graphene\cite{Geim2007,Geim2009}, the ability to isolate atomically thin materials has attracted growing interest within the scientific community owing to a unique set of unprecedented properties suddenly available on a single chip\cite{Wang2012,Chhowalla2013}. Two-dimensional semiconductors are currently enabling conceptually new devices, including light detectors and emitters\cite{Pospischil2016}, transistors\cite{Cho2015} and memories\cite{Tan2015}. The characterization of such devices strongly relies on a set of techniques which combine electrical, optical and spectroscopic measurements. For example, inelastic light scattering (Raman) spectroscopy has been the main tool to study the properties of graphene and its derivatives\cite{Ferrari2013}, while photodetectors based on 2D materials have been largely characterized by Scanning Photocurrent Microscopy (SPCM). To characterize both light-emitting and detecting devices, the knowledge of optical parameters such as the absorption and reflection coefficients of the material is of paramount importance. Commercial instruments are usually used to perform these characterizations in distinct self-standing systems which requires transport of the sample in different environments and retrofitting to accommodate the different holders designed for each specific tool, resulting in the contamination of such sensitive materials.

In this work, we present an experimental apparatus developed to characterize optoelectronic devices based on graphene and other 2D materials. Centred around an upright metallurgical microscope, this system integrates all critical measurements necessary to perform the characterization of an unprecedented range of materials and devices properties in one instrument, with no need to remove the sample from its holder, reducing the risk of contamination or breakage. These measurements include: low frequency electrical transport, SPCM, absorption (transmittance and reflectance), micro-Raman, photoluminescence (PL) and electroluminescence (EL) spectroscopy and mapping, with or without polarized light. A unique aspect of the developed design is the ability to concurrently perform both the electrical and optical measurements, a characteristic that is absent in most common commercial instruments. The system is equipped with multiple laser sources, spanning from UV to red light and two white light sources used for transmission and reflection illumination. Our design delivers the laser light in enclosed paths and an interlock system cuts the laser light to allow access to individual parts, making it extremely safe. High spatial resolution in spectroscopy and SPCM measurements is achieved by diffraction-limited focusing of Gaussian laser beams and a high performance microscope stage. Electrical connections are secured by a custom-built PCB board, designed to reduce electrical noise and allow easy access to devices, without the need for long working distance microscope objectives.

We characterize the instrument with a series of standard measurements: SPCM on graphene-based photodetectors as well as absorption, Raman and PL spectroscopy of a range of 2D materials and organic semiconductors. We prove its high functionality and demonstrate that the achieved resolution, both spatial and spectroscopic, and the overall performance, is equivalent to current commercial technologies or superior in some cases, e.g. in Raman spectroscopy, with the additional benefit of having a compact, multi-purpose, fully customizable, instrument with the potential for installation of additional features.

\section{Operating principles} \label{sec:OperPrinciples}
\subsection{Light emission and detection in atomically thin devices} \label{subsec:Opto01}
Light-matter interaction is at the heart of optoelectronic devices\cite{MarkFox2010Optical,WilsonOptoEl}, which work by converting an electric signal into light or \textit{vice-versa}. For this reason their design requires knowledge of both the optical and electrical characteristics of the materials used. The two main categories in which optoelectronic devices are divided are light emitters (LED) and photodetectors (PDs).
To characterize LEDs, it is necessary to be able to electrically drive the light emission mechanism whilst recording the spectrum and power of the emitted light. The spectral analysis of the emitted radiation is used to gain insight into the physics of the light-emitting material. This is usually achieved by efficiently collecting the emitted light and delivering it to a high-resolution spectrometer with minimal losses or aberrations in the optical path.

The basic experiment performed to characterize a PD consists of shining light onto the device and recording its electrical response, such as measuring the current flow through it or a change in resistance\cite{WilsonOptoEl}. Light can impinge on the whole surface of the device, known as flood illumination, or it can be delivered with a focused laser onto a specific area to allow for a spatially resolved photo-response. In general, the magnitude of the generated photocurrent is given by\cite{WilsonOptoEl}:

\begin{equation}
\label{eq:Photocurrent}
I_{ph} = q\eta_i\frac{P^\beta}{\hbar \nu}(1-\exp^{-\alpha\delta}),
\end{equation}
where $\eta_i = (I_{ph}/q)/\phi_{abs}$ is the internal quantum efficiency of the PD, defined as the number of charges, $q$, collected to produce the photocurrent $I_{ph}$, divided by the number of absorbed photons $\phi_{abs}$, $P_{opt}$ is the incident optical power, $\hbar \nu$ is the energy of the photons, $\alpha$ is the absorption coefficient of the material (see section \ref{subsec:TrRefl01}) and $\delta$ its thickness. The exponent $\beta$ depends on the photocurrent generation mechanism, in general it is equal to $1$ for photovoltaic (PV) devices. A summary of the main quantities used to characterize LEDs and PDs is given in \Cref{tbl:SummryPD}.

The most widely used technique to study PD based on  graphene and TMDs is SPCM, where a laser beam is scanned across the device and the electrical response recorded at each point, producing a two-dimensional map of the photoresponse\cite{Graham2013}. SPCM allows the study of spatially resolved properties of optoelectronic devices, where the spatial resolution is ultimately defined by the laser spot size, $d_s$.

\begin{table}
	\caption{Parameters typically used to characterize optoelectronic devices. Coupling Factors and the wavelength dependence of all quantities is omitted for clarity of notation.}
	\label{tbl:SummryPD}
	\begin{tabular}{ccc}
		\hline
		{\centering Quantity}  & 
		{\centering Symbol} & 
		{\centering Definition\textsuperscript{\emph{a}}}\\
		\hline
		External Quantum Efficiency	&	$\eta_e$, EQE	& $(I_{ph}/q)/\phi_{in}$\\
		Internal Quantum Efficiency	&	$\eta_i$, IQE	& $(I_{ph}/q)/\phi_{abs}$\\
		Wall Plug Efficiency (LED)	&	$\eta$	&	$P_{out}/(I_{in}\cdot V)$\\
		Responsivity	&	$\Re$	& $I_{ph}/P_{opt}$\\
		Specific Detectivity & $D^\star$ & $(S\cdot \Delta f)^{0.5}/\mathrm{NEP}$\\
		\hline
	\end{tabular}
	
	\textsuperscript{\emph{a}} $I_{ph}=$ Measured photocurrent, $\phi_{in}=$ Incident photon flux, $\phi_{abs}=$ Absorbed photon flux, $P_{opt}=$ Incident optical power, $P_{out} = $ Generated optical power, $I_{in}=$ Input electrical power, $V=$ Applied voltage, $S=$ Device area, $\Delta f = $ Operating bandwidth, $\mathrm{NEP}=$ Noise Equivalent Power;
\end{table}

\subsection{Absorption spectroscopy} \label{subsec:TrRefl01}
As shown in \cref{eq:Photocurrent} the absorption coefficient, $\alpha$, of the material has an important role in the response of the device. This is defined as the fraction of power absorbed per unit length into the material, such that the intensity at a distance $x$ in the material is $I(x) = I_0 \exp(-\alpha x)$.

\begin{figure*}
	\includegraphics[scale=1]{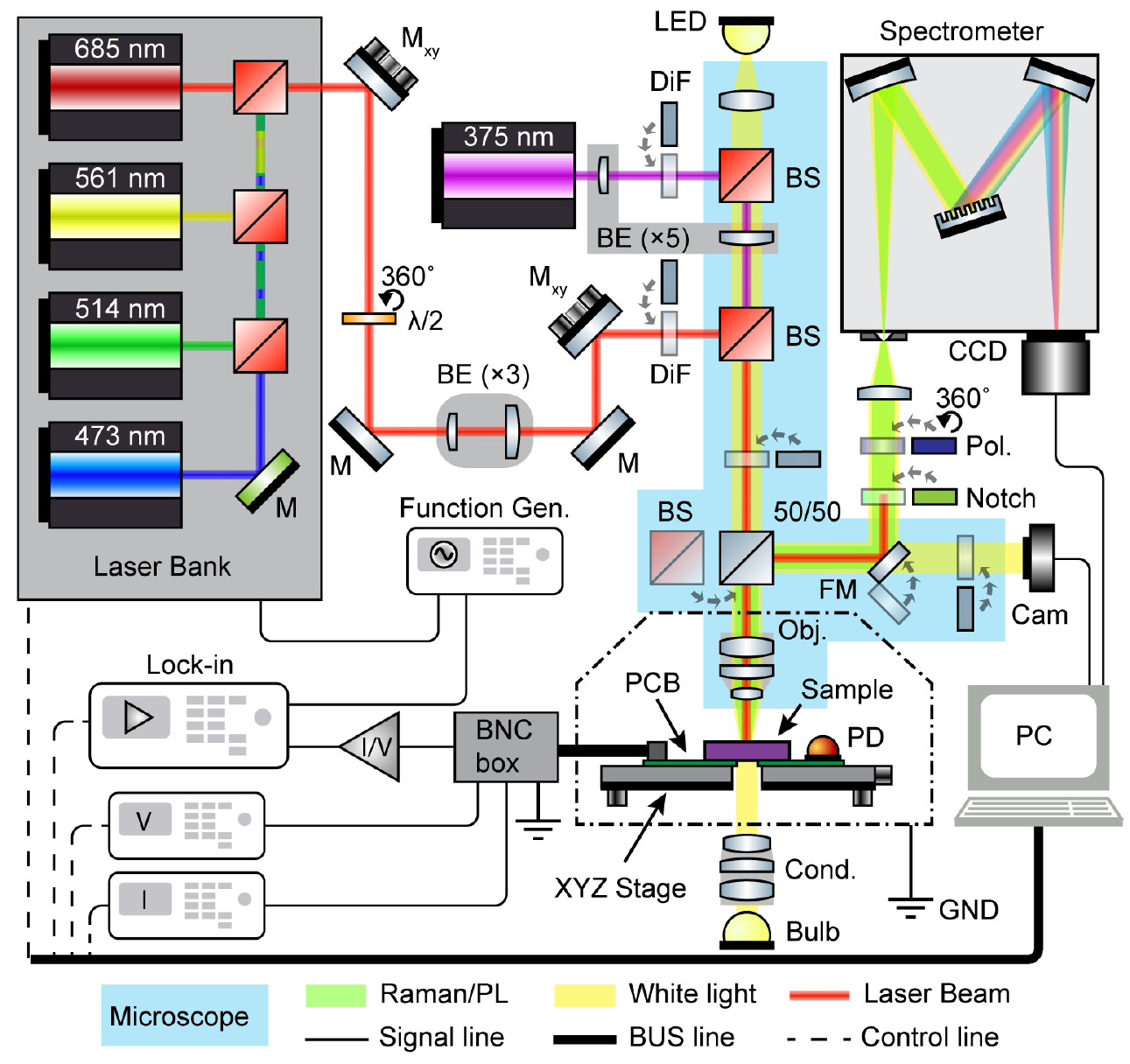}
	\caption{\label{Figure1}\textbf{Schematic of the optoelectronic setup.} Laser light is used for scanning photocurrent mapping, Raman and Photoluminescence (PL) spectroscopy. Four CW diode lasers are fitted in a laser enclosure while the UV laser source is directly attached to the microscope. White light from two different sources can be used for transmittance and reflectance measurements. An XYZ motorized microscope stage allows precise control of the sample position. The dash-dotted line represents the electrically screened light-tight enclosure of the sample stage. Signal lines carry electric/data signals to be measured, control lines carry the signals to configure the instruments, the lasers and the other sources, BUS line comprises USB and GPIB. Abbreviations: mirror (M), kinematic mirror (M$_\mathrm{xy}$), half-wavelength plate ($\lambda/2$), beam expander (BE, followed by magnification), drop-in filter (DiF), beam splitter (BS, dichroic in red), Polariser/Analyser (Pol.), white light (WL), voltage (V) or current (I) sources/meters, flip mirror (FM), sample holder (PCB), photodetector (PD), condenser (Cond), microscope objective (Obj), imaging camera (Cam), spectroscopy camera (CCD), ground line (GND). Created with ComponentLibrary\cite{ComponentLibray} symbols.}
\end{figure*}

For a thin film deposited on a thick substrate, as with TMDs and other 2D crystals, $\alpha$ is given by\cite{Swanepoel1983}:

\begin{equation}
\label{eq:Methods_AbsorptionCoefficient_SWAN1983}
\alpha = -\frac{1}{\delta} \ln \left( \frac{P + \sqrt{P^2+2 Q T \left( 1- R_2 R_3 \right)}}{Q} \right),
\end{equation}
where:

\begin{subequations}
	\begin{align}
	\label{eq:AbsorptionCoefficient_Q}
	&Q = 2T(R_1 R_2 + R_1 R_3 - 2 R_1 R_2 R_3) ,
	\\
	\label{eq:AbsorptionCoefficient_S}
	&P = (R_1 - 1)(R_2 - 1)(R_3 - 1) ,
	\\
	\label{eq:AbsorptionCoefficient_R1}
	&R_1 = \left[\frac{1-n}{1+n}\right]^2, R_2 = \left[\frac{n-s}{n+s}\right]^2,	R_3 = \left[\frac{s-1}{s+1}\right]^2.
	\end{align}
\end{subequations}
Here $n$ and $s$ are the refractive indices of the medium and substrate respectively, $R_1$ is the reflectance of the air/medium interface, $R_2$ is the reflectance of the substrate/medium interface and $R_3$ the reflectance of the substrate/air interface. Each $R_i$ can be either measured directly or computed from the knowledge of $n$ and $s$.

Absorption spectroscopy is a viable method to determine the bandgap of a semiconductor and its nature, direct or indirect. A plot of $\alpha$ has a specific shape\cite{MarkFox2010Optical,Terashima1987} for energies close to the bandgap: for indirect transitions (neglecting excitonic effects) $\alpha \propto \left( E_{ph} - E_g \pm \hbar \Omega \right)^m$, where $E_{ph}$ is the incident photon energy, $E_g$ is the bandgap energy, $\hbar \Omega$ is the associated phonon energy and $m$ is equal to $2$ for allowed and $3$ for forbidden transitions; for direct transitions $\alpha \propto \left( E_{ph} - E_g \right)^{0.5}$. In 2D materials it has been shown that the bandgap of few-layer graphene can be tuned with an applied electric field\cite{Khodkov2015} and in TMDs the bandgap energy changes with the number of layers and induced strain\cite{Roldan2015}. TMDs also show a transition of the bandgap from indirect to direct when they are thinned to single-layers\cite{Wang2012}. This large variety of physical phenomena makes absorption spectroscopy an important tool for studying these novel materials.

\subsection{Inelastic light scattering and luminescence spectroscopy} \label{subsec:Raman01}
The inelastic scattering of light, or Raman effect, involves the scattering of a photon by a phonon in the examined material\cite{FerraroRamanBook}. Each phonon mode can be represented by the Raman tensor $\mathrm{\mathbf{R_i}}$, for which basis vectors are the crystallographic axes of the crystal (or molecule). The intensity of the Raman signal is given by:

\begin{equation}
\label{eq:RamanIntensityPol}
I_R = c_1 \sum_{i} \int \left\| \overrightarrow{e_S} \cdot \mathrm{\mathbf{R_i}} \cdot \overrightarrow{e_I}^\mathrm{T} \right\|^2 d\Omega_S,
\end{equation}
where the integration is carried over the collection angle of the microscope objective $d\Omega_S$, $\overrightarrow{e_I}$ represents the excitation and $\overrightarrow{e_S}$ the scattered photons respectively, $c_1$ is a material dependent factor and $\mathrm{T}$ represents the transpose of the vector. Thus, by changing the excitation and collection axes and the polarization of the light it is possible to determine the crystallographic orientation of the material and identify the different vibrational modes\cite{Kolb1990}. 

The intensity of the Raman scattered light is dependent on the local temperature $T$ and the ratio between Stokes ($I_S$) and anti-Stokes ($I_{aS}$) intensities can be used as a temperature probe\cite{Hart1970}, using the well-known expression\cite{Jellison1983}:

\begin{equation}
\label{eq:Stokes-antiStokes}
\frac{I_S}{I_{aS}} = F\cdot C_{ex}\cdot \exp\left(\frac{h \omega_i}{k_B T}\right),
\end{equation}
where $F$ is a parameter that depends on the optical constants of the material and $C_{ex}$ is a coefficient which accounts for the efficiencies of the optics and gratings at different wavelengths.

\begin{figure*}{}
	\includegraphics[scale=1]{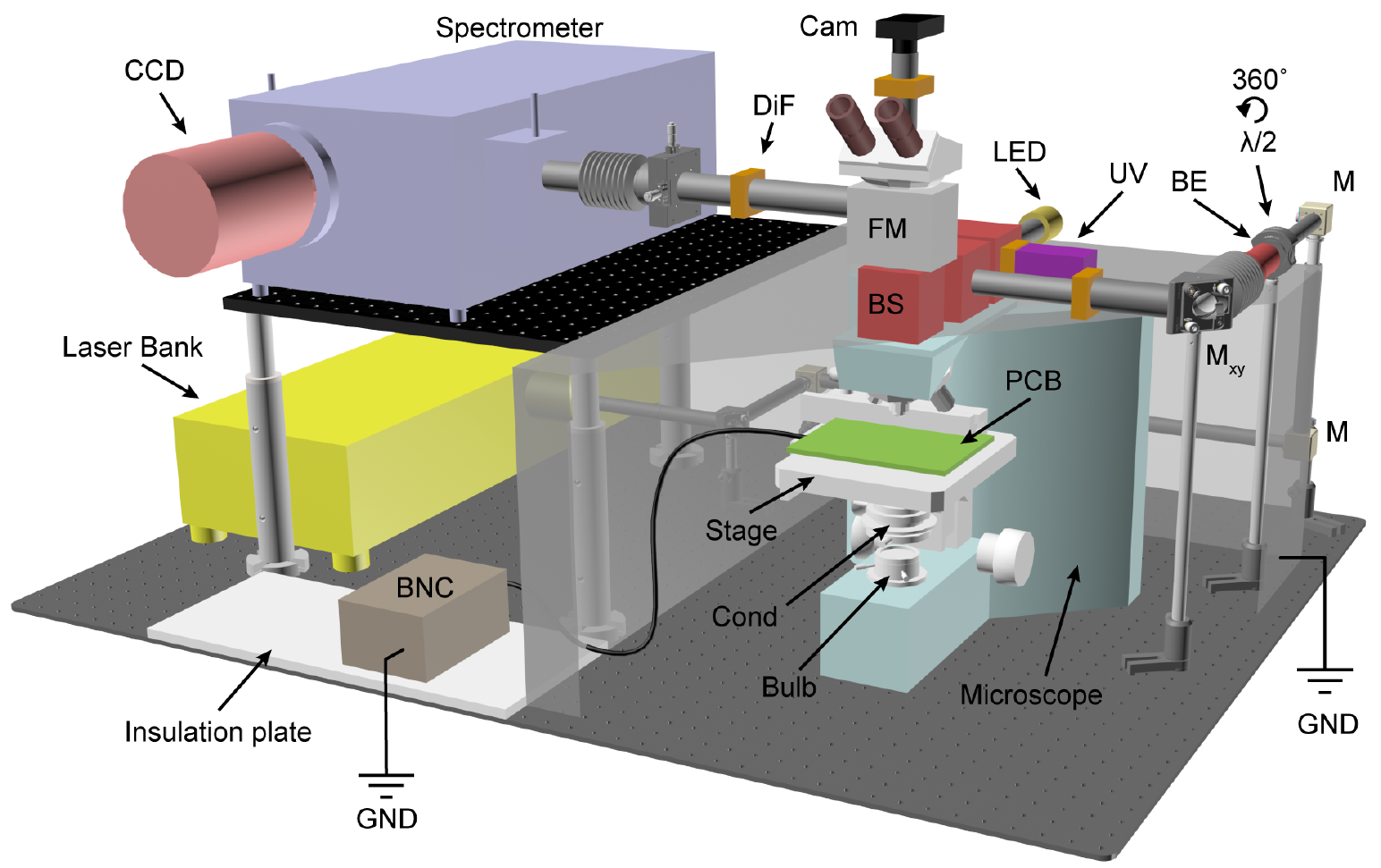}
	\caption{\label{Figure2}\textbf{3D model showing the implementation of the setup.} The system is built on a vibration-isolated optical table ($120\times90$ cm$^2$). Four main components can be identified: upright microscope with custom-built EPI illumination system and motorized XYZ stage; Custom-built laser-enclosure; Spectrometer and CCD camera; Sample-holder PCB board with BNC break-out box. The electrically screened, light-tight, enclosure of the sample is shown in transparency and sectioned, this is directly grounded. Tube lenses are used to cover the light path of the laser beams, magnetic interlocks are located in the screen enclosure and in the BS cubes, effectively making the system a Class 1 laser product. Labels and abbreviations as in Figure \ref{Figure1}.}
\end{figure*}

Closely related to Raman spectroscopy are luminescence phenomena, where radiative transitions of electrons from an excited state to a lower level lead to the emission of light. These transitions can be initiated by the absorption of a photon, known as Photoluminescence (PL), fluorescence and phosphorescence, or by the application of an electric field through the material, known as Electroluminescence (EL). In both cases, detailed measurement of the spectral intensity of the emitted light can give insight into the properties of the material and the performance of the examined device. In a typical PL experiment the excitation source is a laser, with a wavelength appropriate for exciting resonant electronic transitions.

\section{Instrumentation} \label{sec:Intrumentation}
\subsection{Optics} \label{subsec:Instr_Optics}
\Cref{Figure1} shows a schematic of the optical and electrical components of the system whilst a 3D model presenting the actual implementation of the components is shown in \cref{Figure2}. A set of visible wavelength CW diode lasers (\textit{Coherent} OBIS 375LX, 473LS, 514LX, 561LS and \textit{Omicron} LuxX 685, with powers ranging from $30$ to $50$ mW) is fitted in an enclosure system, which facilitates the alignment of each laser and can host up to 6 different wavelengths. Each laser is digitally modulated and the power is adjusted using an analog signal. Continuous modulation and TTL operation can be achieved by use of a digitally-controlled signal generator. Careful choice of dichroic mirrors inside the laser enclosure allows multiple wavelengths to be used at the same time. The laser light is then delivered with a series of kinematic and fixed mirrors into the illumination path of the optical microscope (\textit{Olympus} BX51), see blue shaded area in \cref{Figure1}, after a $\times 3$ beam expansion. A dichroic beam splitter (BS) is used to direct the light in the custom built epi-illumination section of the microscope. A rotatable (360$\degree$) achromatic $\lambda/2$ waveplate (\textit{ThorLabs} AHWP05M-600) is used to rotate the polarization of the laser light. A custom-built drop-in-filter (DiF) system is used to introduce neutral density (ND), polarisers, notch and band-pass filters in the optical path of the lasers and the microscope: the same filter holders and housings are used throughout the system, ensuring full compatibility in each section. Invisible (UV) laser light is fed into the microscopy directly from the epi-illumination BS cubes after a $\times 3$ beam expansion to secure a superior laser beam resolution delivered to the sample.

\begin{figure}{}
	\includegraphics[scale=1]{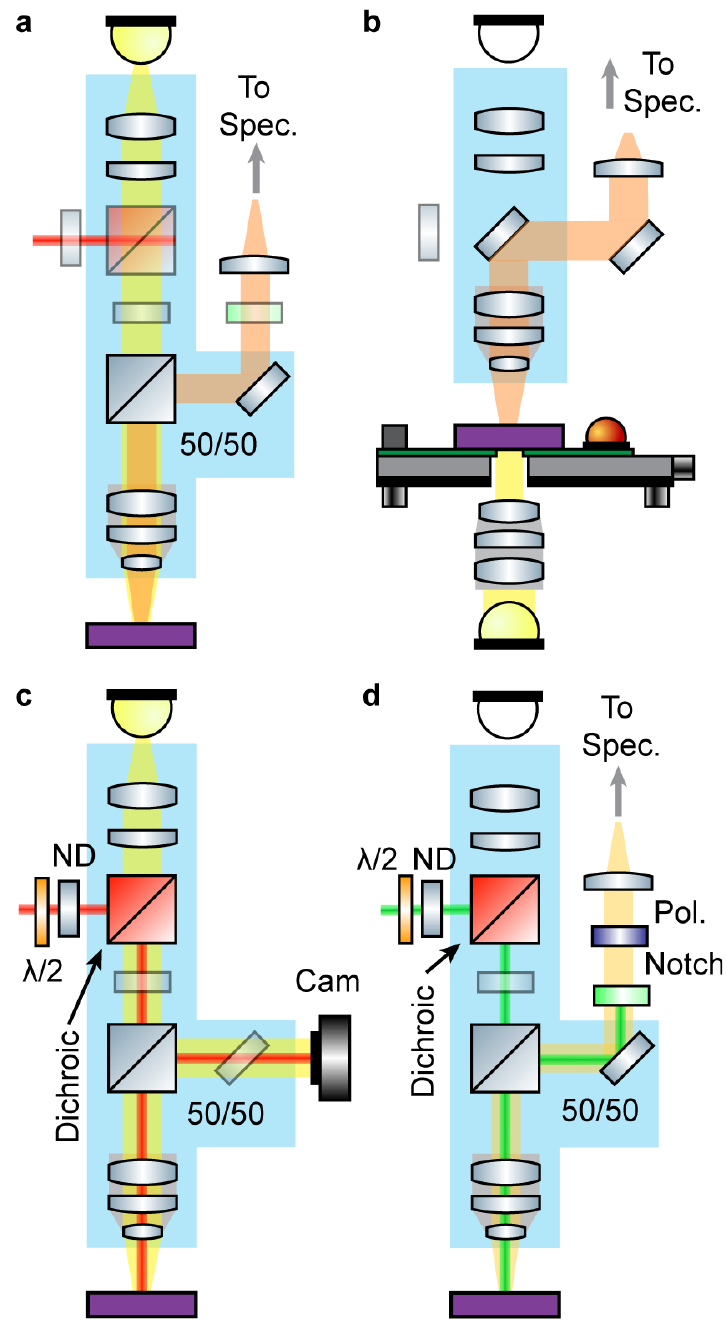}
	\caption{\label{Figure3} \textbf{Flexible Optical path configurations.} \textbf{a}, Reflectivity and \textbf{b}, Transmittance measurements using white light. \textbf{c}, Scanning photocurrent mapping with laser light; \textbf{d}, Raman/Fluorescence/Photoluminescence spectroscopy, including polarization. \textbf{c} and \textbf{d} can be combined and performed at the same time. UV laser delivery system is omitted. Symbols and abbreviations as in Figure \ref{Figure1}.}
\end{figure}

White light for reflection microscopy is provided by a white LED along the optical path of the laser beams, while transmittance illumination is achieved using the microscope's built-in lamp and condenser lens. Both the white and laser light are directed to the objective using either a $50\%/50\%$ BS or the appropriate dichroic band-pass (BP) mirror. The reflected light collected by the objective is directed to a CCD imaging camera or, by using a flip mirror (FM), to the entrance slit of the spectrometer. The microscope is fitted with $Olympus$ MPLanFL-N Semi-Apochromat infinity-corrected lenses with $\times 5$, $\times 10$, $\times 20$ and $\times 50$ magnification.

The spectroscopy section is composed by a focusing achromatic lens ($ThorLabs$ AC254-150-A) with x/y and focus adjustment. The spectrometer is a \textit{Princeton Instruments} Acton SP2500, equipped with three dispersion gratings ($1200$ g/mm with $500$ nm and $750$ nm blaze, and $1800$ g/mm with $500$ nm blaze) and a \textit{Princeton Instruments} PIXIS400-eXcelon back-illuminated, peltier cooled, CCD camera. In our multi-purpose instrument, each grating is used for different measurements as they differ by number of grooves and blaze wavelength (at which the efficiency is at maximum and the same for both S and P polarizations). In white light spectroscopy applications, corrections to account for the optics and gratings efficiencies, at different wavelengths, can be applied using \textit{Princeton Instruments} IntelliCal system.

The microscope stage is a \textit{Prior Scientific} OptiScan ES111 with ProScan III controller with a minimum step size is $10$ nm. Focus control is also achieved through the same controller.

The whole system is built on a vibration-isolated $120\times90\,\mathrm{cm}^2$ optical table. The laser light delivery system is enclosed within \textit{ThorLabs} stackable tube lens system, allowing a light-tight connection. The microscope stage, foot and objective turret is covered with a light-tight enclosure, formed by a metal frame and a conductive fabric curtain connected to the ground of the electrical circuit to ensure shielding from electro-magnetic noise. The front of the curtain can be lifted to allow access to the stage and it is fitted with magnetic interlocks. The light-tight delivery system and enclosure, together with the magnetic interlocks, make the system a Class $1$ laser product. The same light-tight tubes are used also to deliver light to the spectrometer, enabling the use of our instrument without the need of a darkened room.

The custom-built epi-illumination system allows flexible and quick configuration of the microscope for different measurements. As shown in \cref{Figure3}, the optical path can be configured for Raman, PL, fluorescence and transmission/reflection spectroscopy and laser light illumination (as in SPCM) simply by replacing or removing the appropriate filters and BS. Each BS unit is fitted with magnetic interlocks for safety purposes. \Cref{Figure3}a shows the configuration  for reflectance measurements: light from the white LED travels freely (no laser BS fitted) after being collimated and reaches the objective though a $50\%/50\%$ BS; reflected light is collected by the same objective and partially transmitted by the BS to a FM which directs it into the spectrometer. A similar arrangement is used for transmittance spectroscopy, as shown in \cref{Figure3}b. In this case the BS is replaced with a mirror and the white light from the incandescent bulb is collimated by the microscope condenser and, after travelling in the sample, collected by the objective and directed to the spectrometer. \Cref{Figure3}c shows the configuration for laser-light illumination and SPCM: in this case a dichroic BS is used to direct the laser light into the objective and the reflected light is partially transmitted to the imaging camera for direct assessment and focussing; the LED can be used at the same time for imaging. The configuration for Raman and luminescence spectroscopy is shown in \cref{Figure3}d: in this case the scattered light is collected by the objective and directed to the spectrometer via the FM, a notch filter is used to remove the Rayleigh component while ND and laser-line filters can be used to attenuate the beam power and clean it from spurious wavelengths. The latter two configurations can be used at the same time to perform Raman or PL mapping together with SPCM. Polarized Raman spectroscopy can be performed by interposing a polarizing filter (analyser) between the notch filter and the spectrometer, while the laser polarization is controlled by the $\lambda/2$ waveplate.

A photodetector (PD) mounted on the stage, see \cref{Figure1}, is used to automatically measure the power of the laser light after a measurement. The ability to modulate the power of the lasers allows the system to span a power range (with the $\times50$ objective) between $\sim 2$ nW and $\sim20$ mW in steps of $0.4\cdot10^{-OD}$ mW, where $OD$ is the optical density.

In our setup solid-state diode lasers are used and all the optical components are chosen so that to minimize deviations from the TEM$_{00}$ laser mode, which has a Gaussian intensity distribution. As the beam is transmitted through a circular aperture of radius $r_a$, such as the pupil aperture of the objective lens, the transferred power is reduced by a factor $(1-\exp(-2/T^2))$, where $T=r_b/r_a$ is defined as the Gaussian beam truncation ratio and $r_b$ is the radius of the beam measured at the point at which the beam intensity falls by $50\%$ (FWHM). The focused beam profile is Gaussian for $T<0.5$ and converges to the Airy pattern for $T \rightarrow \infty$. The focused spot size diameter ($d_s$) can be expressed as\cite{Urey2004, Self1983}:

\begin{equation}
\label{eq:Methods_SpotSize}
d_s = \frac{K\lambda}{2 \mathrm{NA}},
\end{equation}
where $\mathrm{NA}$ is the numerical aperture of the lens, $\lambda$ is the wavelength of the laser and $K$ is the $k$-factor for truncated Gaussian beams, which is a function of $T$ only. \textcite{Urey2004} computed approximate expressions for $K$ for the two cases $T<0.5$ (Gaussian) and $T>0.4$ (Truncated Gaussian). In the same work an expression to determine the depth of focus, $\Delta z$ (i.e. the distance along the optical axis at which the irradiance drops by 50$\%$), was determined:

\begin{equation}
\label{eq:Methods_DepthFocus}
\Delta z = K_2 \lambda f_\#^2,
\end{equation}
where $f_\#$ is the ratio between the clear aperture diameter and the focal length of the lens (also called $f$-number) and $K_2$ is the focus $k$-factor, again a function of $T$ only. In our apparatus we found that $T = 1.03$ for the UV ($375$ nm) laser and $T = 0.52$ for the visible lasers when using the $\times 50$ objective. In table \ref{tbl:Methods_SpotSize} we show the results obtained for our system using equations \ref{eq:Methods_SpotSize}, \ref{eq:Methods_DepthFocus} and the results in reference \onlinecite{Urey2004}. The calculated values show that our system is able to focus the laser light well in the diffraction-limit of the objective lens, allowing ultra-high spatial resolution. The very narrow depth of focus allows substrate contributions to be minimized.

\begin{figure}{}
	\includegraphics[scale=0.9]{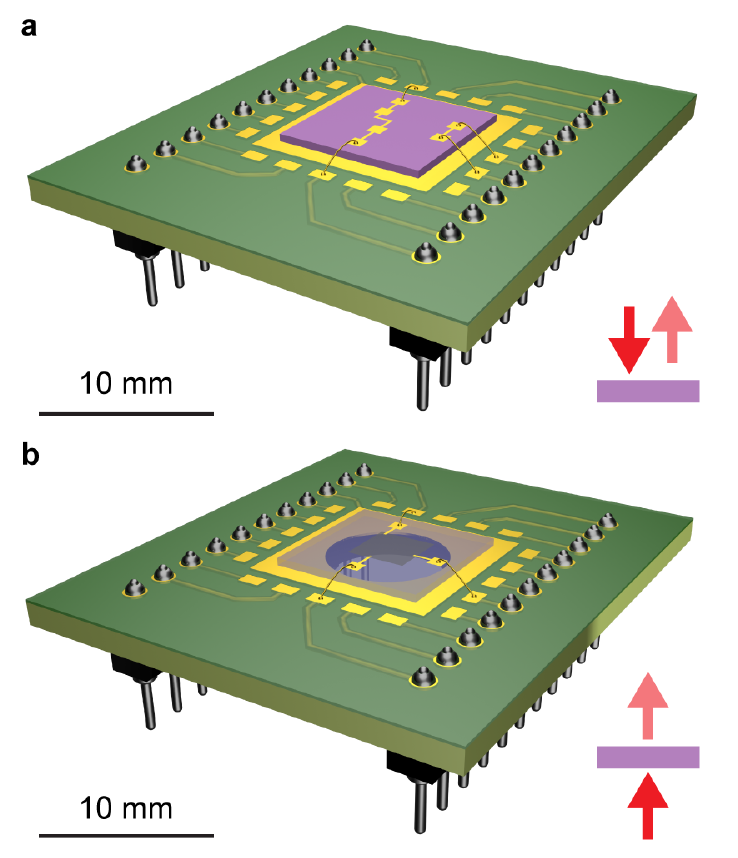}
	\caption{\label{Figure4} \textbf{Custom PCB chip-carrier boards.} \textbf{a}, Opaque substrates configuration where reflected and laser light is used. \textbf{b}, Transparent substrate configuration where transmitted light is used. Yellow areas are gold-coated pads. Scale bars are $10$ mm.}
\end{figure}

\begin{table}
	\caption{Laser spot diameters and depth of focus of the lasers in our apparatus with the $\times50$ objective.}
	\label{tbl:Methods_SpotSize}
	\begin{tabular}{ccc}
		\hline
		$\lambda$ (nm)  & $d_s$ (nm) & $\Delta z$ (nm)\\
		\hline
		375	&	264	& 158\\
		473	&	445	& 268\\
		514	&	484	& 291\\
		561 &	528	& 320\\
		685	&	645	& 390\\
		\hline
	\end{tabular}
\end{table}

\subsection{Electronics} \label{subsec:Instr_Electr}
Optoelectronic devices based on graphene and 2D materials are usually fabricated from thin flakes deposited on a Si substrate capped with a SiO$_2$ layer, where electrical contacts are defined via lithography and metal evaporation. We designed a chip carrier for our microscope which allows versatile and easy connection of such devices. As shown in \cref{Figure4}, the carrier is composed of a PCB board ($34\times29\,\mathrm{mm}^2$) with two standard $11$-way pin strips, where $20$ of these pins are connected to gold-coated pads and one to a central $15\times15\,\mathrm{mm}^2$ gold-coated pad (one is not connected). The central pad is used to contact, using silver paint, the doped silicon substrate which provides electrostatic gating. The other pads are used to connect the device's pads to the PCB board using wire bonding. The design of the chip carrier differs from standard, commercially available, boards since the sample is not buried in a plastic or ceramic case, it is, instead, above the PCB and the soldering pads. In this way the objective lens cannot come into contact with parts of the chip carrier and long-working distance lenses are not required, thus improving the maximum achievable spatial resolution and minimum laser spot size. The chip carriers can also be fabricated with a central aperture, as shown in \cref{Figure4}b, allowing transmitted light to be used to characterize devices fabricated on transparent substrates.

Thanks to the standard single-in-line package (SIP) pins, the chip carriers are connected to female sockets on a second PCB board which is anchored on the microscope stage. This board is then connected though a shielded multi-core cable to a breakout BNC box, which enables connections with different measuring instruments, such as lock-in amplifiers, voltage/current amplifiers and multimeters, as shown in figures \ref{Figure1} and \ref{Figure2}. In order to screen the devices from external electric fields, the light-tight enclosure surrounding the stage is made of conductive fabric and connected to the ground of the circuit, together with the walls of the BNC box. All measuring instruments are decoupled from the rest of the electric instruments \textit{via} insulating transformers, while signal and bus lines are kept isolated from the measuring PC \textit{via} opto-electronic decouplers (\textit{National Instruments} GPIB-120B).

Full automation of the system is achieved with a custom-made LabView-based software which is able to communicate to all electronic instruments via GPIB/USB bus and modulate the lasers \textit{via} digital-to-analog interface (DAC) (\textit{National Instruments} NI-DAQ); Native software is used to control the spectrometer and CCD camera, interfaced with the same LabView software.

\section{Performance of multi-purpose microscope system} \label{sec:Characterization}
\subsection{Optoelectronic devices} \label{subsec:Opto02}
\begin{figure}{}
	\includegraphics[scale=1]{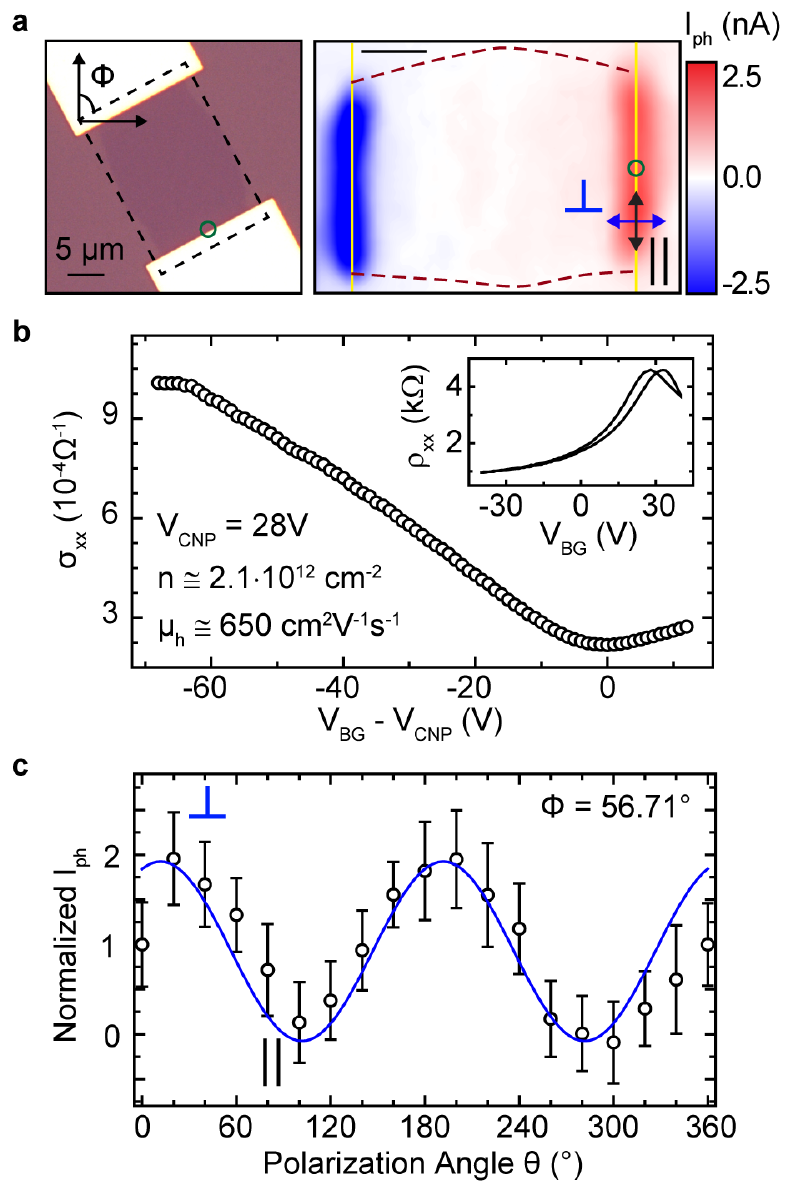}
	\caption{\label{Figure5} \textbf{Characterization of a graphene-based PD.} \textbf{a}, Optical micrograph of the single-layer graphene device in FET configuration (left) and SPCM map at $\lambda_{\mathrm{exc}} = 473$ nm acquired with our microscope (right); black-dashed box marks the mapped area, red-dashed lines mark the graphene flake boundary. Scale bars are $5\,\mu$m. \textbf{b}, Electrical characterization of the graphene FET, acquired in the same setup, showing the conductivity as a function of back-gate voltage. \textbf{c}, Polarization dependence of the observed photocurrent $I_{ph}$ at the graphene-metal contact (green circle in panel \textbf{a}), maximum photocurrent is observed for light polarized $\perp$ to the metal contact. $\phi$ marks the angle between the vertical (y) axis of the microscope stage ($\theta = 0\degree$) and the graphene-metal interface, solid blue line is the expected dependence (see main text).}
\end{figure}

As an example of the ability of our setup to perform characterization of optoelectronic devices, we present typical measurements performed on a graphene-based photodetector. The apparatus is setup as in \cref{Figure3}c, where the WL source is used for imaging purposes and switched off during the measurements. The device is comprised of a single-layer exfoliated graphene flake, contacted in a two-terminal geometry with Cr/Au contacts, on a Si substrate capped with SiO$_2$ gate oxide, see \cref{Figure5}a, left panel. Details on the fabrication and properties of such devices are beyond the scope of this paper and discussed elsewhere in the literature\cite{Koppens2014,Woessner2016,Graham2013a,Tielrooij2015}. In \cref{Figure5}a, right panel, we show the SPCM map of such device acquired with a $473$ nm laser ($24$ kW/cm$^2$ incident power) in steps of $0.5\,\mu$m, under a source-drain bias $V_{SD} = 10$ mV and back-gate voltage $V_{BG} = 0$ V. The map shows strong photocurrent at the graphene/metal interface, in agreement with previous reports\cite{Koppens2014,Tielrooij2015}. We then characterize the electronic response of the device directly on the microscope stage. \Cref{Figure5}b shows the back-gate sweep (typical for a graphene FET) of the device under the same $V_{SD}$: from this we can extrapolate several parameters, such as the level of doping (holes in this case) $n \simeq 2.1\cdot10^{12}$ cm$^2$ and the field-effect hole mobility $\mu_h \simeq 650$ cm$^2$V$^{-1}$s$^{-1}$. All measurements are performed using a \textit{DL} Model 1211 current amplifier, an \textit{Ametek} 7270 DSP lock-in amplifier and a \textit{Keithley} 2400 SourceMeter to provide the source and gate biases.

\begin{figure}
	\includegraphics[scale=1]{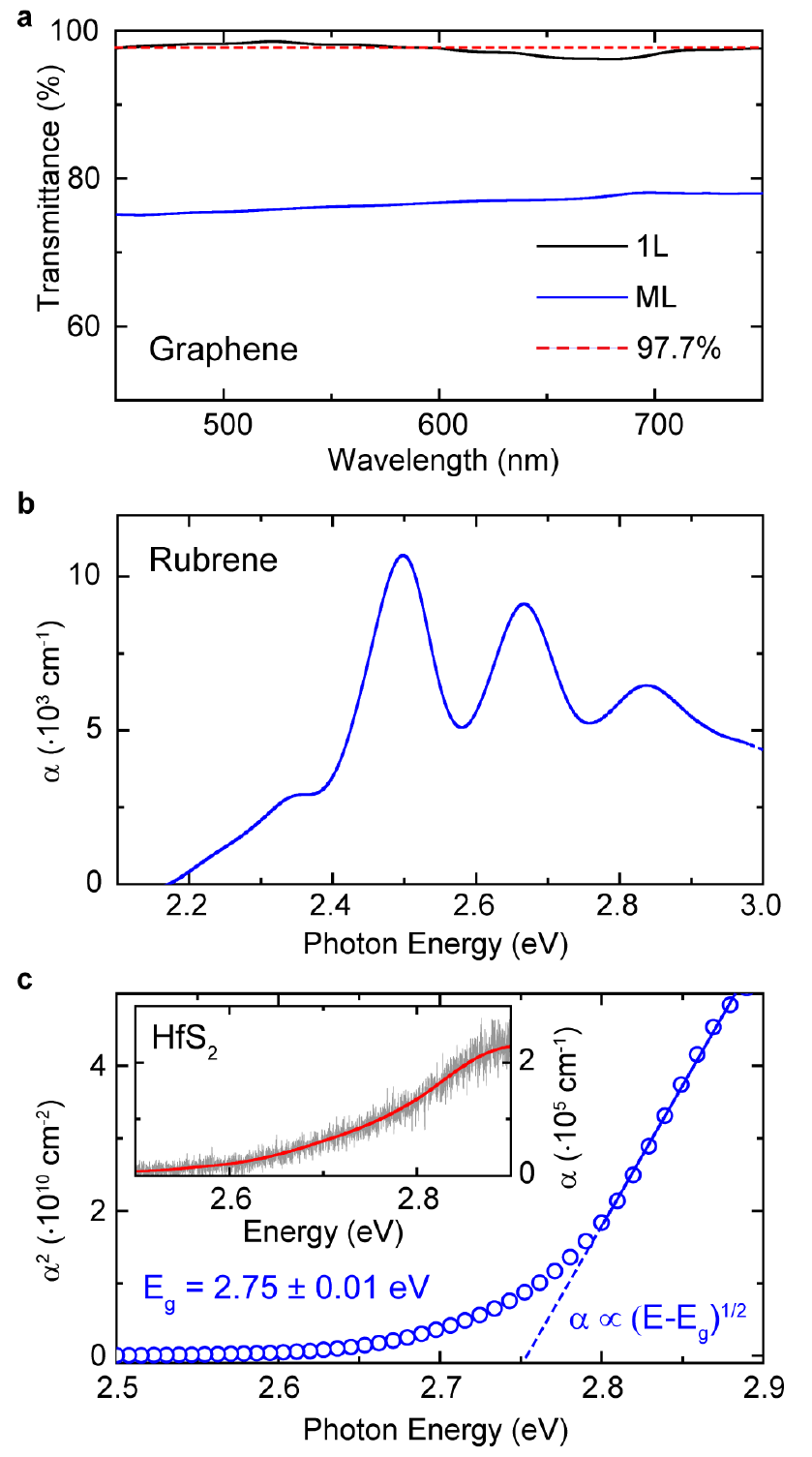}
	\caption{\label{Figure6} \textbf{Absorption spectroscopy characterization.} \textbf{a}, Visible transmittance of CVD-grown single- (1L) and multi-layer (ML) graphene, dashed line marks the theoretical $97.7\%$ transmittance of single-layer graphene; \textbf{b} Unpolarized absorption coefficient $\alpha$ of a Rubrene crystal, measured perpendicular to the ab facet; \textbf{c}, Absorption coefficient $\alpha$ of ultra-thin ($\sim 25$ nm) HfS$_2$ and direct-gap determination, $E_g = 2.75\pm0.01$ eV.}
\end{figure}

As a further system characterization, we demonstrate the polarization capabilities of our apparatus by measuring the polarization dependence of the observed photocurrent. We focus the laser at the graphene/metal interface and rotate the $\lambda/2$ waveplate, in order to change the polarization of the incident light, while recording the photocurrent. It is well known that, in this type of device, hot-carriers dynamics in graphene, combined with the direct absorption of the metal contacts, lead to a polarization-dependent photocurrent\cite{Tielrooij2015}. In \cref{Figure5}c we show a plot of the photocurrent as a function of polarization angle $\theta$. We observe that $I_{ph}$ is maximum when the polarization is orthogonal ($\perp$) to the metal contact and minimum when it is parallel ($\parallel$) to it, as expected\cite{Tielrooij2015}. The angle $\theta$ is defined as the angle between the polarization of the laser and the vertical (y) axis of the microscope stage. Since our device sits at an angle $\phi \simeq 57\degree$ with respect to y (see \cref{Figure5}a), the polarization dependence has a ``phase shift'' of the same amount, i.e. $I_{ph} \propto \sin(2(\theta-\phi))$, as verified in \cref{Figure5}c (solid blue line).

The measurements presented here are in general not possible on a commercial instrument. The use, for example, of a commercial Raman setup is common for such characterisation, however, the lack of customizability, control over laser power and reliable electrical contacts represents a challenge for the characterisation of small flakes.

\subsection{Absorption spectroscopy} \label{subsec:TrRefl02}
In \cref{subsec:TrRefl01} we showed that, in order to extrapolate the absorption coefficient of a material, we need to perform transmittance and reflectance measurements. These can be done in our system by using the optical configurations shown in \cref{Figure3}a-b. As a characterization of the performance of our setup we present a series of measurements of transmittance and absorption coefficient for different materials. 

\begin{figure*}
	\includegraphics[scale=1]{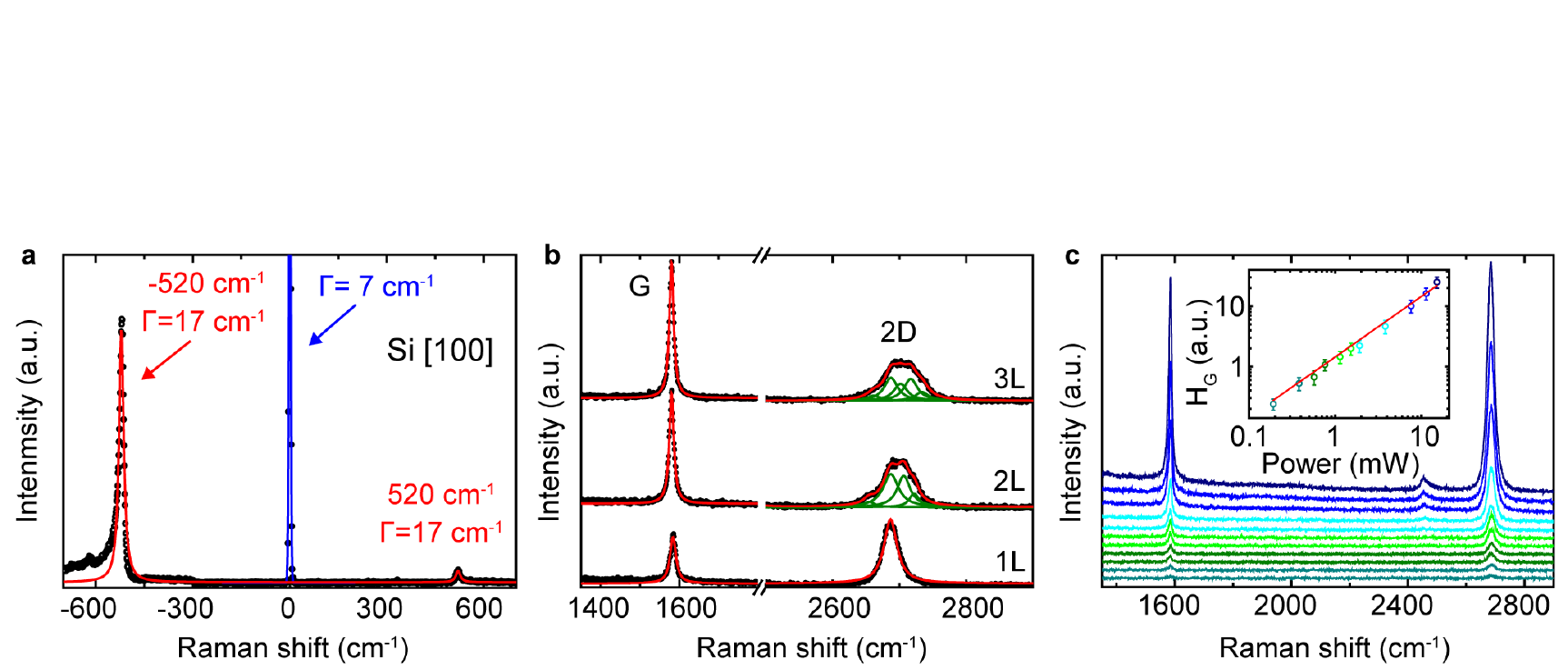}
	\caption{\label{Figure7} \textbf{Raman spectroscopy characterization.} \textbf{a}, Stokes and anti-Stokes bands of Si [100] at room temperature; \textbf{b}, Resonant Raman spectra of single- (1L), bi- (2L) and tri- (3L) layer graphene, red solid lines are cumulative fit to Lorentzian curves (shown in green); \textbf{c}, Raman spectrum of single-layer graphene as a function of incident optical power and power-dependence of the G-peak height (inset), solid line marks a slope of $1$, curves shifted for clarity. All spectra are acquired with $\lambda_{\mathrm{exc}}=514$ nm at $2.1$ MW/cm$^2$ incident power and $2$ s acquisition time.}
\end{figure*}

\Cref{Figure6}a shows the visible transmittance of single-layer and multi-layer CVD-grown graphene. As expected\cite{Mak2008}, single layer graphene shows a flat transmittance $\sim 97.7\%$ across the whole visible spectrum, while nickel-grown multi-layer CVD graphene\cite{Bointon2015} shows a transmittance of $\sim 75-80\%$. The contribution of the transparent substrate, quartz in this case, has been subtracted from the data by performing a calibration measurements on a bare substrate. The absorption coefficient of two materials is determined b measuring transmittance and reflectance and applying \cref{eq:Methods_AbsorptionCoefficient_SWAN1983}. The first is a thin layer of organic crystalline semiconductor, Rubrene. The optical and electrical properties of this organic semiconductor have been intensively studied\cite{Irkhin2012}. Therefore, it offers a good standard to test the performance of the developed multi-purpose microscope system. \Cref{Figure6}b shows the measured $\alpha(\lambda)$ of a Rubrene crystal on glass, using unpolarized light, perpendicular to the a-b facets and parallel to the c facet of the crystal (see Ref. \onlinecite{Irkhin2012} for details). The measured spectrum is in very good agreement with literature, both in the intensity and position of the peaks\cite{Irkhin2012}. The second material is ultra-thin HfS$_2$. This is a well known TMD which was well characterized in the past as a bulk crystal\cite{Greenaway1965} and recently as a thin flake on a substrate. \Cref{Figure6}c, inset, shows the absorption coefficient of a $\sim 25$ nm thick flake of freshly-exfoliated HfS$_2$. The values agree well with literature\cite{Greenaway1965,Terashima1987}. Owing to the proportionality between $\alpha$ and $E$ (see \cref{subsec:TrRefl01}), we are able to extrapolate the bandgap of the material by plotting $\alpha^2$ as a function of $E$. The intercept of the extrapolated linear part mark the direct bandgap of the material $E_g = 2.75\pm0.01$ eV, in very good agreement with established values\cite{Greenaway1965}.

Absorption spectroscopy can be performed on the same sample-holder as the electrical measurements (see \cref{Figure4}), a feature that is not present in commercial instruments. This gives the ability to relate the optical properties to the electrical response of the material directly \textit{in situ}.

\subsection{Raman spectroscopy} \label{subsec:Raman02}
To demonstrate the ability of our set-up to also perform inelastic light spectroscopy we characterized it by studying two well-known materials: Si and graphene. \Cref{Figure7}a shows the Raman spectrum of commercial-grade Si, acquired with $\lambda_{\mathrm{exc}} = 514$ nm at $1.0$ MW/cm$^2$ incident power for $10$ s, where the first-order transverse-optical phonon (1TO) mode mode at $\sim 520\,\mathrm{cm}^{-1}$ is captured in both Stokes (negative shift) and anti-Stokes (positive shift) regime in a single snapshot. The FWHM ($\Gamma$) of the Rayleigh line ($0\,\mathrm{cm}^{-1}$) is only $7\,\mathrm{cm}^{-1}$, and the Lorentzian fit of the Si modes gives $\Gamma = 17\,\mathrm{cm}^{-1}$. Using \cref{eq:Stokes-antiStokes} for $T = 291$ K and $F=0.97$ (see Ref. \onlinecite{Jellison1983} for details), we find an expected value of $I_S/I_{aS} = 11.92$; the experimental value extrapolated from \cref{Figure7}a is $I_S/I_{aS} = 22 \pm 2$, giving an experimental correction factor $C_{ex} \simeq 1.84$. In the following discussion we will adopt the common convention for which the Stokes lines are reported with positive shift, instead of negative.

The resonant Raman spectrum, acquired with our setup, of exfoliated graphene is shown in \cref{Figure7}b. We observe the G peak at $\sim 1580\,\mathrm{cm}^{-1}$, originating from the $E_{2g}$ mode and the double-resonant 2D band at $\sim 2680\,\mathrm{cm}^{-1}$, originating from the $A_{1g}$ mode. The 2D band is sensitive to the number of layers: it is given by a single Lorentzian peak for single-layer graphene, the convolution of four Lorentzian peaks for bi-layer graphene and the convolution of six peaks for tri-layer graphene\cite{Malard2009}. In \cref{Figure7}b we show the ability of our system to resolve such multi-peak structures (green lines) and thus discriminate the number of layers in graphene flakes. \Cref{Figure7}c shows the same spectrum of single-layer graphene acquired at different incident powers. The inset shows the height of the G peak as a function of incident power, adhering to the expected linear relationship. This serves to confirm that no artefacts are introduced in the spectra by the experimental apparatus. It is worth noting that we are able to resolve clearly the G and 2D bands with an incident power density as low as $50$ kW/$\mathrm{cm}^2$ with an acquisition time of only $2$ s for the whole spectral range. For comparison, a commercial Raman spectrometer requires an acquisition time $>10$ s with an incident laser power $>600$ kW/$\mathrm{cm}^2$ to obtain the same counts from single-layer graphene.

\begin{figure}
	\includegraphics[scale=1]{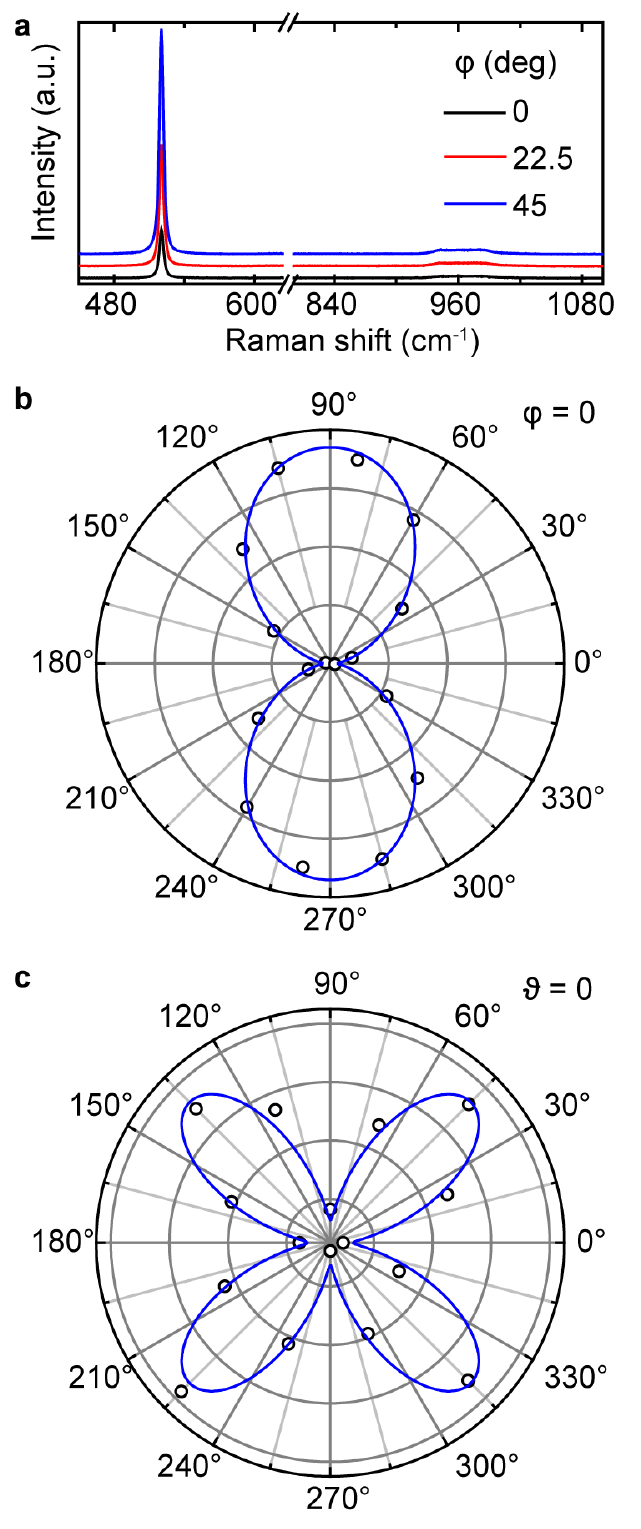}
	\caption{\label{Figure8} \textbf{Polarized Raman spectra of Si [100].} \textbf{a}, Raman spectra of Si showing the first-order (TO)-phonon mode of Si at $\sim 520\,\mathrm{cm}^{-1}$ and two-phonon mode at $\sim 900-1000\,\mathrm{cm}^{-1}$. \textbf{b}, Polar plot of the height of the (TO)-phonon mode as a function of the angle between polarization directions $\vartheta$ ($\varphi = 0$) and \textbf{c}, same plot as a function the rotation angle about the [100] axis $\varphi$ ($\vartheta = 0$), solid line is given by \cref{eq:Raman_SiliconPol}. $\lambda_{\mathrm{exc}}=514$ nm, $P_{\mathrm{opt}} = 1.2$ MW/cm$^2$, $2$ s acquisition time.}
\end{figure}

In our setup it is possible to perform polarized Raman spectroscopy using the configuration shown in \cref{Figure3}d. This technique has recently been applied to the investigation of graphene\cite{Yoon2008} and TMDs\cite{Chenet2015}, where strong polarization dependence is given by valley anisotropy, highlighting the role of Raman spectroscopy in the study of electron-phonon coupling. To demonstrate the polarization capabilities of our setup, we characterized the dependence of the Raman modes of Si upon incident polarization. Figure \ref{Figure8}a shows the Raman spectrum of Si with the [100]-surface perpendicular to both the incident and the scattered light ($z xx \overline{z}$ and $z xy \overline{z}$ configurations), acquired with $\lambda_{\mathrm{exc}} = 514$ nm at $2.1$ MW/cm$^2$ incident power for $2$ s. We observe the 1TO mode at $520\,\mathrm{cm}^{-1}$ and a two-phonon mode at $\sim 900-1000\,\mathrm{cm}^{-1}$. Using \cref{eq:RamanIntensityPol} and the expression for the Raman tensor of the 1TO mode, we find that the Raman intensity of this mode is given by\cite{Kolb1990}:

\begin{equation}
\label{eq:Raman_SiliconPol}
I_R (\vartheta, \varphi) = c_2 \sin^2(\vartheta - 2\varphi),
\end{equation}
where $\vartheta$ is the polarization angle between the incident and scattered light. As the angle between the incident polarization and the [100] axis $\varphi$ is varied, we observe a decrease of the 1TO peak, while the two-phonon mode changes only slightly, as shown in \cref{Figure8}a. \Cref{eq:Raman_SiliconPol} is verified in \cref{Figure8}b-c where the intensity of the 1TO mode is plotted as a function of $\vartheta$ and $\varphi$, respectively. The angle $\varphi$ is varied by rotating the polarization of the laser with a $\lambda/2$ waveplate, while the angle $\vartheta$ is defined by the angle of the analyser with respect to the waveplate (see \cref{Figure4}d). In this way there is no need to rotate the sample, making it possible to have electrical connections to the devices. At the same time, this technique can be used to assess the orientation of the sample, since a measurement of the $\vartheta$ dependence will give a value of $\varphi \neq 0$ if the [100] axis is not aligned with the incident polarization angle.  

\subsection{Luminescence spectroscopy} \label{subsec:PLEL02}
\begin{figure}{}
	\includegraphics[scale=1]{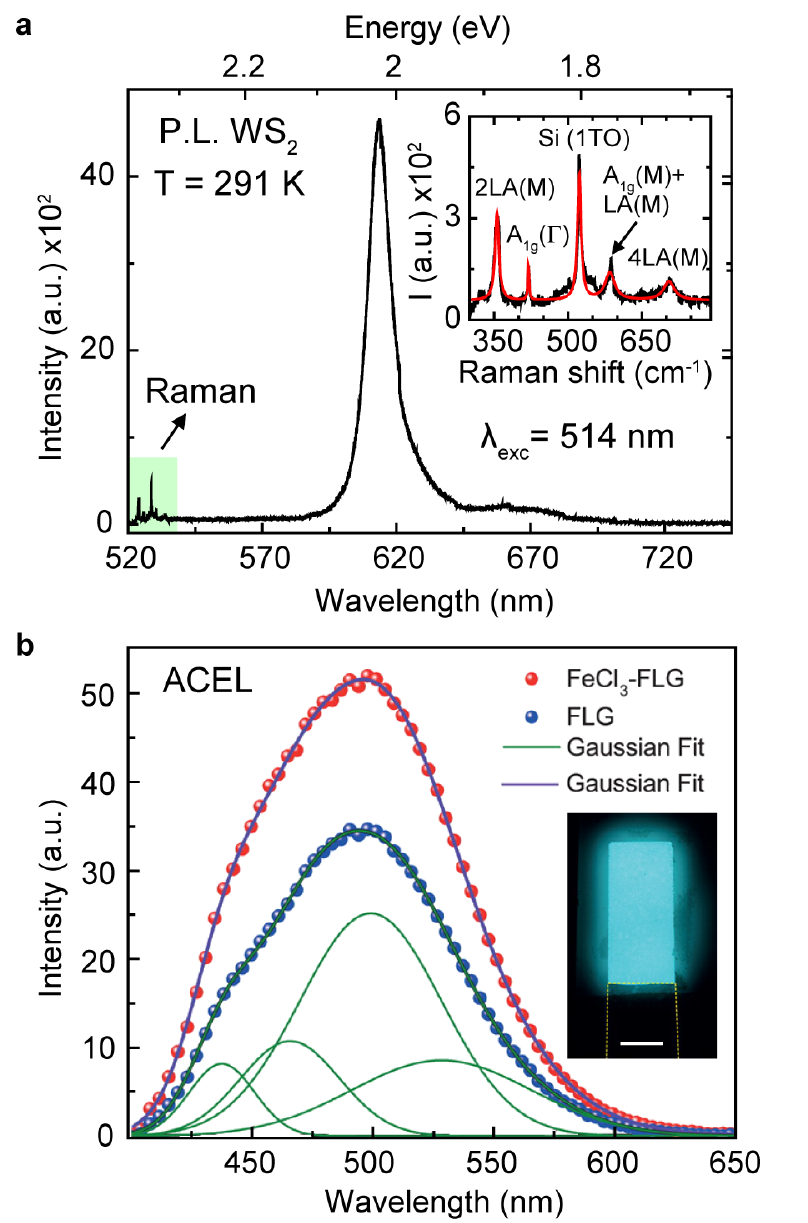}
	\caption{\label{Figure9} \textbf{PL spectroscopy and EL devices characterization.} \textbf{a}, Room temperature Photoluminescence of CVD-grown single-layer WS$_2$, $\lambda_{\mathrm{exc}}=514$ nm, $P = 5.1$ kW/cm$^2$, $10$ s acquisition time. Inset: zoom on observed Raman modes. \textbf{b}, ZnS$_2$-based ACEL device spectral emission: comparison between few-layer graphene and FeCl$_3$-intercalated graphene electrodes. Inset: picture of the working device. Scale bar is $1$ cm. Reprinted with permission from Elias Torres Alonso et al. Homogeneously Bright, Flexible, and Foldable Lighting Devices with Functionalized Graphene Electrodes. ACS Applied Materials \& Interfaces 2016, 8, 16541. Copyright 2016 American Chemical Society.}
\end{figure}

The ability of our system to measure PL phenomena is shown using single-layer WS$_2$, a semiconducting TMD with good potential in optoelectronic applications\cite{Zeng2016,Yao2016}. The room-temperature PL spectrum is shown in \cref{Figure9}a. This is acquired using the configuration shown in \cref{Figure3}d, with $\lambda_{\mathrm{exc}} = 514$ nm, $P = 5.1$ kW/cm$^2$ and $10$ s acquisition time; the power is kept low to avoid causing damage to the sample. A strong peak is observed, centred at $2.02$ eV and the Raman modes can be also observed in the same spectrum (highlighted in the green box). Fit with Lorentzian curves of the Raman modes (inset) gives the following values: $2LA(M) = 353\,\mathrm{cm}^{-1}$, $A_{1g} (\Gamma) = 420\,\mathrm{cm}^{-1}$, $A_{1g}(M)+LA = 585\,\mathrm{cm}^{-1}$ and $4LA(M) = 705\,\mathrm{cm}^{-1}$. Both the PL and Raman spectra are in perfect agreement with commonly accepted values\cite{Elias2013}. The use of the intensity calibration provided by the spectrometer allows full integration of light-scattering spectroscopy, as it is well exemplified by the above results. The Raman and PL spectra are, in fact, corrected in intensity to take into account the transmittance of all filters and optical parts, as well as the efficiencies of the grating and of the CCD. Such feature is usually not found in commercially-available Raman spectrometers.

As an example of EL device characterization we use a ZnS$_2$-based ACEL device with both graphene and functionalized graphene transparent electrodes\cite{Alonso2016}. Figure \ref{Figure9}b shows the spectral emission of such device with graphene and FeCl$_3$-intercalated graphene electrodes. The spectra can be decomposed into multiple Gaussian peaks, corresponding to the intraband transitions\cite{Manzoor2003} in ZnS$_2$. The use of our instrument allows the accurate comparison of the two electrode materials and the performance of the different devices (see \textcite{Alonso2016} for details).

Fluorescence spectroscopy can be also performed using the configuration shown in \cref{Figure3}d, by replacing the notch filter with a low-pass using the UV laser (or other suitable wavelength) as excitation source.

\section{Conclusion} \label{sec:Conclusions}
We have presented the design of a new multi-purpose instrument for the characterization of optoelectronic devices based on 2D materials, capable of performing multiple electrical and optical measurements simultaneously. We demonstrated the performance of the setup with a series of techniques, involving a multitude of materials, showing results at the level of the state-of-the-art commercial equipment. The ability to perform low-frequency electrical transport measurements, SPCM, Raman, absorption and PL spectroscopy, combined with full automation, high sensitivity and low noise, make our instrument ideal to overcome the challenges imposed by the advent of atomically-thin materials in optoelectronic devices.

Customization of each section allows multiple routes for future upgrades and expansion, these will include: a vacuum chamber microscope stage with temperature control, multi-wavelength Raman spectroscopy, auto-focusing and high-frequency (RF) measurements.

The cost and size is very contained, making it suitable for small laboratories and the safety features introduced make it very easy to operate, with minimal training required. The use of commercial parts gives the advantage of being replicated easily, following the emerging open-hardware approach to develop novel scientific instruments which is attracting growing attention by the scientific community\cite{OpenHardwareNature} as a highly efficient means of innovation in fast growing research fields such as 2D materials\cite{Novoselov2012_Roadmap,Ren2014}. To this end, our open-hardware system will be of high interest to the industrial community since it demonstrates that the integration of critical measurements for the characterization of a wide range of optoelectronic applications and devices based on graphene and 2D materials will not require complex or bespoke instruments. The scientific community will benefit from such instrument as the aforementioned integration allows rapid sample screening, reducing the possibility of contamination and therefore increasing the research throughput.

\begin{acknowledgments}
The authors wish to thank Ana I. S. Neves for providing the graphene samples; Elias Torres Alonso and George Karkera for providing data on the ACEL devices; Jake Mehew for useful discussion;  Paul Wilkins and Adam Woodgate for workshop assistance. 
S. Russo and M.F. Craciun acknowledge financial support from EPSRC (Grant no. EP/J000396/1, EP/K017160/1, EP/K010050/1, EP/G036101/1, EP/M001024/1, EP/M002438/1), from Royal Society international Exchanges Scheme 2016/R1, from European Commission (FP7-ICT-2013-613024-GRASP), from The Leverhulme trust (grant title "Quantum Drums" and "Room temperature quantum electronics") and from DSTL (grant scheme Sensing and Navigation using quantum 2.0 technologies and UK-France bilateral scheme).
\end{acknowledgments}

\section*{Author contributions}
All authors contributed to the design of the system. A.D.S. assembled the instrument, performed the characterization measurements, prepared the schematic figures and wrote the manuscript, with contributions from all authors. G.F.J. and N.J.T. designed the PCB sample-holders and provided advice on the design. G.F.J. provided the Rubrene and WS$_2$ samples and analysed the absorption measurements. M.F.C. and S.R. supervised the project and provided advice on the measurements and the manuscript.

\section*{Competing financial interests}
The authors declare no competing financial interests.

\bibliography{references_paperG16_AdeS}

\end{document}